\documentclass[%
reprint,
superscriptaddress,
 amsmath,amssymb,
 aps,
floatfix,
]{revtex4-1}

\usepackage{dcolumn}
\usepackage{bm}

\usepackage[dvipdfmx]{graphicx,color}
\usepackage{color}

\usepackage{ulem} 

\usepackage[dvipdfmx,
colorlinks=true, citecolor=blue, urlcolor=blue, linkcolor=blue,
setpagesize=false, bookmarks=false
]{hyperref}

\begin{document}
\title{Spin loop-current textures in Hubbard models}
\author{Kazuya Shinjo}
\affiliation{Computational Quantum Matter Research Team, RIKEN Center for Emergent Matter Science (CEMS), Wako, Saitama 351-0198, Japan}
\author{Shigetoshi Sota}
\affiliation{Computational Materials Science Research Team,
RIKEN Center for Computational Science (R-CCS), Kobe, Hyogo 650-0047, Japan}
\affiliation{Quantum Computational Science Research Team,
RIKEN Center for Quantum Computing (RQC), Wako, Saitama 351-0198, Japan}
\author{Seiji Yunoki}
\affiliation{Computational Quantum Matter Research Team, RIKEN Center for Emergent Matter Science (CEMS), Wako, Saitama 351-0198, Japan}
\affiliation{Computational Materials Science Research Team,
RIKEN Center for Computational Science (R-CCS), Kobe, Hyogo 650-0047, Japan}
\affiliation{Quantum Computational Science Research Team,
RIKEN Center for Quantum Computing (RQC), Wako, Saitama 351-0198, Japan}
\affiliation{Computational Condensed Matter Physics Laboratory,
RIKEN Cluster for Pioneering Research (CPR), Saitama 351-0198, Japan}
\author{Takami Tohyama}
\affiliation{Department of Applied Physics, Tokyo University of Science, Tokyo 125-8585, Japan}

\date{\today}
             
 
\begin{abstract}
The recent experimental observations of loop current in Sr$_{2}$IrO$_{4}$, YBa$_{2}$Cu$_{3}$O$_{7}$, and Sr$_{14}$Cu$_{24}$O$_{41}$ have inspired a theoretical study that broadly redefines loop current as a manifestation of quantum liquid crystals.
Using the density-matrix renormalization group method, we investigate the emergence of spin loop-current (sLC) textures in carrier-doped (i) excitonic insulators, (ii) orbital-selective Mott insulators, and (iii) two-dimensional Mott insulators, modeled by a two-orbital Hubbard model on a ladder lattice in (i) and (ii) and a single-orbital Hubbard model on a square lattice in (iii).
Calculating the spatial distribution of spin current around a bond to which a pinning field is applied, we find conditions for longer-ranged sLC correlations.
In system (i), when using the model parameters employed to describe the excitonic condensation, we find that a sLC texture appears near half filling, associated with an excitonic condensation in a spin channel.
In system (ii), using typical sets of model parameters for BaFe$_{2}$Se$_{3}$, we find that a sLC texture appears at electron fillings where a block-type antiferromagnetism develops.
In system (iii), introducing a next-nearest-neighbor hopping $t'\sim -0.25$ (in units of the nearest-neighbor hopping) suggested for high-$T_\text{c}$ cuprates, we find that an axial-sLC texture emerges at hole-carrier density $\delta=0.125$, where the charge stripe simultaneously appears.
\end{abstract}
\maketitle


%
\section{Introduction}\label{sec1}
Unexpected phenomena often emerge in quantum many-body systems, and they are most pronounced when strong quantum fluctuations are present in strongly correlated electron systems.
According to Landau, many-body phases of matter exhibit spontaneous symmetry breaking at low temperatures, and phase transitions are characterized by symmetry changes.
However, for example, quantum spin liquids do not break any symmetry and hence they are not characterized by any local order parameter, indicating the presence of quantum phases beyond the description of Landau's symmetry-breaking theory.
Furthermore, an intermediate state between spontaneous symmetry-broken and symmetry-unbroken states has been proposed as a quantum liquid crystal~\cite{Kivelson1998, Kivelson2003, Vojta2009, Fradkin2010, Fradkin2012, Fradkin2015}.
Namely, a quantum liquid crystal is regarded as a quantum state with partially broken spatial symmetry, and it can exhibit unconventional properties since the partial symmetry-breaking order can interplay with other intrinsic orders such as superconducting and magnetic orders.

A quantum nematic state~\cite{Yamase2000, Halboth2000, Oganesyan2001, Metzner2003, Kee2003, Kee2008} is one of the most well-known examples of quantum liquid crystals, but there is another quantum liquid crystal with spontaneous loop current.
Although quantum states with loop current, also called staggered-flux or orbital-antiferromagnetic states, have a long history in the field of strongly correlated electron systems~\cite{Halperin1968, Affleck1988, Marston1989, Nersesyan1989, Schulz1989}, they have attracted renewed interest in the past decade.
This is due to the recent discovery of a series of quantum states with various charge loop-current (cLC) textures in Sr$_{2}$IrO$_{4}$~\cite{Jeong2017, Maruyama2021}, YBa$_{2}$Cu$_{3}$O$_{7}$~\cite{Mangin-Thro2015, Zhao2017}, and Sr$_{14}$Cu$_{24}$O$_{41}$~\cite{Bounoua2020} via the improved measurements of the Kerr effect, polarized neutron scattering, magnetic torque, and second-harmonic generation.
Theoretically, cLC textures have been extensively studied based on the Hubbard models~\cite{Scalapino2001, Greiter2007, Thomale2008, Nishimoto2009, Kung2014} as a cLC long-range order~\cite{Varma1997, Chakravarty2001, Varma2005a} or its fluctuation~\cite{Wen1996} suggested to characterize the pseudogap phase in high-$T_\text{c}$ cuprate superconductors.
However, no numerical evidence of cLC textures has been reported in single- and three-orbital Hubbard models~\cite{Scalapino2001, Greiter2007, Thomale2008, Nishimoto2009, Kung2014}.
It has been proposed that quantum states with cLC textures can be present in a generalized Hubbard ladder~\cite{Marston2002, Schollwock2003, Wu2003, Fjaerestad2006} and spinless Hubbard model~\cite{Sur2018, JuliaFarre2020, JuliaFarre2023} if unrealistic interactions for real materials are introduced.

The recent experimental observations described above have encouraged further theoretical investigation of loop-current textures from a different point of view.
One such viewpoint is to explore the possibility of quantum states with spin current rather than charge current~\cite{Kunes2016, Kontani2021}.
Due to the difficulty of treating strongly correlated electron systems, many theoretical studies rely on the mean-field or perturbation theory. 
Here, instead, we address the question of whether non-perturbative treatment can elucidate quantum states with spin loop-current (sLC) textures in the Hubbard models.

For this purpose, we employ the density-matrix renormalization group (DMRG) method to investigate the possibility of quantum states with sLC in the Hubbard models.
In particular, we consider three kinds of strongly correlated electron systems in this paper: carrier-doped (i) excitonic insulators, (ii) orbital-selective Mott insulators, and (iii) two-dimensional Mott insulators, modeled by a two-orbital Hubbard ladder with a crystal field for (i) and (ii) and a single-orbital Hubbard model on a square lattice for (iii).
In the two-orbital model, we introduce the Hund coupling and pair hopping in addition to the Coulomb interaction.
Calculating the spatial distribution of spin current around a bond to which a pinning field is applied, we examine whether there are cases in which sLC correlations can be enhanced.

We find that such cases exist, i.e., sLC textures emerging for each of these systems (i)--(iii).
In system (i), the sLC correlations are enhanced, if we introduce interorbital hopping in the model.
This is consistent with a previous study of dynamical mean-field theory~\cite{Kunes2016}.
System (ii) is known to exhibit stripe- and block-type antiferromagnetic (AFM) phases~\cite{Herbrych2019}, which have been observed experimentally in BaFe$_2$$X_{3}$ ($X=$ Se, S)~\cite{Caron2012, Mourigal2015, Chi2016}.
We find a quantum state with sLC textures in the vicinity of the block-type AFM phase.
In contrast to system (i), the interorbital hopping is not relevant to generating sLC textures here. 
The carrier density where the sLC correlations are significantly enhanced corresponds to that in a generalized Kondo-Heisenberg model (GKHM) where vector chirality order appears~\cite{Sroda2021}.
In system (iii), we find that the sLC correlations are enhanced when we introduce an appropriate value of the next-nearest-neighbor hopping $t'$.
The sLC textures found here emerge at hole density $\delta=0.125$, and they coexist with charge stripes having a spatial modulation period of 4.

Our findings clearly show that sLC textures can spontaneously emerge by introducing carriers and/or orbital degrees of freedom to increase quantum fluctuations, which induce quantum liquid crystallinity.
Therefore, our approach, which does not rely on the mean-field or perturbation theory, will ad another perspective to the study of quantum liquid crystals.

The rest of this paper is organized as follows. 
We first introduce the phenomenological theory of a quantum liquid crystal with sLC textures in Sec.~\ref{sec2}.
We then show the results of our DMRG study of two-orbital Hubbard ladders in Sec.~\ref{sec3}.
In Sec.~\ref{sec3a}, we numerically demonstrate that sLC textures emerge in a carrier-doped excitonic insulator.
Here, the interorbital hopping and crystal field are necessary for inducing the excitonic condensation with sLC textures.
In Sec.~\ref{sec3b}, we show that sLC textures arise in a carrier-doped orbital-selective Mott insulator, where the introduction of an appropriate carrier density and the different degree of localization in the two orbitals are both necessary to achieve sLC textures.
Next, we show in Sec.~\ref{sec4} the results on a single-orbital Hubbard model on a square lattice, for which the intermediate value of $t'$ is necessary for sLC textures that emerge at hole density $\delta=0.125$.
In Sec.~\ref{sec5}, we summarize this paper.
In Appendix~\ref{app-a}, we supplement the explanation of the pinning-field approach used in our DMRG study with further numerical results. 
In Appendix~\ref{app-b}, we demonstrate numerically that the spontaneous hybridization indeed occurs in the carrier-doped two-orbital Hubbard model on a ladder lattice, suggesting an excitonic condensation. 
In Appendix~\ref{app-c}, we provide additional analysis on the spin current induced by a pinning field, which exhibits power-law decay as a function of distance from the bond at which a pinning field is applied.

\section{Quantum liquid crystal with spin loop-current textures}\label{sec2}

Quantum liquid crystals in two dimensions have been discussed on the basis of the Pomeranchuk instability~\cite{Pomeranchuk1958} in Landau's Fermi liquid theory~\cite{Baym, Leggett}.
Considering a charge channel, quantum liquid crystals can be described with phase separation, charge nematic~\cite{Oganesyan2001},  and cLC. Here, we focus on a spin channel, which might yield sLC.
In the momentum space, the order parameters for quantum liquid crystals in a spin channel are given as
\begin{align}
\langle Q_{l,x}^{a} \rangle =&\sum_{\bm{k},\tau,\tau'} \langle \psi_{\tau} ^{\dag}(\bm{k}) \sigma^{a}_{\tau \tau'} \psi_{\tau'}(\bm{k}) \rangle \cos (l\theta_{\bm{k}}) \label{eq:Q1} 
\end{align}
and 
\begin{align}
\langle Q_{l,y}^{a} \rangle =&\sum_{\bm{k},\tau,\tau'} \langle \psi_{\tau} ^{\dag}(\bm{k}) \sigma^{a}_{\tau \tau'} \psi_{\tau'}(\bm{k}) \rangle \sin (l\theta_{\bm{k}})
\label{eq:Q2}
\end{align}
along the $x$ and $y$ directions, respectively~\cite{Wu2004, Wu2007}, where $\psi_{\tau}(\bm{k})$ is an annihilation operator for an electron with momentum $\bm{k}$ and spin $\tau=\uparrow,\downarrow$, $\sigma^{a}_{\tau \tau'}$ indicates the $(\tau,\tau')$ element of the Pauli matrix $\sigma^{a}$ with $a=x$, $y$, and $z$, and $\theta_{\bm{k}}$ is the azimuthal angle of $\bm{k}$. 
$l$ denotes an orbital angular momentum and thus it is a non-negative integer (for quantum liquid crystals, $l>0$).

A quantum liquid crystal with nonzero $\langle Q_{l,x/y}^{a}\rangle$ might be induced by the Pomeranchuk instability of the Fermi surfaces~\cite{Pomeranchuk1958}.
The Landau parameters $F_{l}$ quantify the strength of the forward scattering interactions among quasiparticles at low energies close to the Fermi surface in a spin channel.
The thermodynamic stability of the Fermi liquid state requires that the Landau parameters $F_{l}$ not be too negative.  
Namely, the thermodynamic instability occurs when $F_{l} < -(2l+1)$.
The most typical Pomeranchuk instabilities are found in the $s$-wave channel: the Stoner ferromagnetism with $\langle Q_{0,x/y}^{a}\rangle\ne0$ is induced at $F_{0}<-1$.
For the $p$-wave channel, $\langle Q_{1,x/y}^{a}\rangle \neq 0$ represents spin currents flowing along the $x/y$ direction, leading to spin-dipole moments in momentum space.
For $l\geq 1$~\cite{Wu2004, Wu2007}, $\langle Q_{l,x/y}^{a}\rangle$ breaks spin-orbital symmetry as originally proposed in $^{3}$He~\cite{Leggett1975, Leggett}.
An emergent Rashba~\cite{Rashba1960}-Dresselhaus~\cite{Dresselhus1955}-like spin-orbit coupling can lead to a quantum state with  sLC textures~\cite{Wu2004, Wu2007, Raghu2008}.
In the following sections, we show the numerical results of our microscopic study for quantum states with sLC textures.

\section{Two-orbital Hubbard ladders}\label{sec3}

\subsection{Carrier-doped excitonic insulators}\label{sec3a}

The order parameter $\langle Q_{l,x/y}^{a} \rangle$ of a quantum liquid crystal is a particle-hole pair condensation, which is analogous to an exciton condensation in semimetals~\cite{Mott1961, Knox1963, Keldysh1965, Halperin1968, Halperin1968b}.
The emergence of sLC textures associated with an exciton condensation has been suggested previously in (dynamical) mean-field analysis~\cite{Kunes2016, Geffroy2018, Nishida2019}.
The sLC textures are associated with an exciton condensation in the spin channel, which is stabilized in the presence of Hund's coupling and pair hopping~\cite{Kaneko2014}.
Furthermore, the pair field of an exciton condensation can be imaginary when interorbital hopping is suitably selected~\cite{Kunes2014, Kunes2016}.

Here, we numerically investigate sLC textures associated with an exciton condensation in the spin channel.
Our study is based on the two-orbital Hubbard model on a ladder lattice.
In the SU(2)-symmetric form, it is given by the following Hamiltonian: 
\begin{align}\label{eq-twoHub}
\mathcal{H}_\text{TH}=&-\sum_{\langle i,j \rangle,\gamma,\gamma',\tau} t_{\gamma \gamma'} (c_{i,\gamma,\tau}^{\dag}c_{j,\gamma',\tau} + \text{H.c.})\nonumber \\
&+U\sum_{i,\gamma}n_{i,\gamma,\uparrow}n_{i,\gamma,\downarrow} 
+ (U-5J_\text{H}/2)\sum_{i}n_{i,a} n_{i,b}\nonumber \\
&-2J_\text{H}\sum_{i}\bm{S}_{i,a}\cdot \bm{S}_{i,b}
+J_\text{H}\sum_{i}(P_{i,a}^{\dag}P_{i,b}+\text{H.c.}),
\end{align}
where $c_{i,\gamma,\tau}^{\dag}$ is the electron creation operator at site $i$ with spin $\tau\,(=\uparrow,\downarrow)$ and orbital $\gamma\,(=a,b)$.
$n_{i,\gamma}=\sum_{\tau}n_{i,\gamma,\tau}$ is the total charge density operator for orbital $\gamma$ at site $i$, where $n_{i,\gamma,\tau} = c_{i,\gamma,\tau}^{\dag} c_{i,\gamma,\tau}$.

The first term of the Hamiltonian $\mathcal{H}_{\rm TH}$ represents the nearest-neighbor electron hopping from orbital $\gamma$ at site $i$ to orbital $\gamma'$ at site $j$ and vice versa with a hopping amplitude $t_{\gamma \gamma'}$.
$\langle i,j\rangle$ is a nearest-neighbor pair of sites $i$ and $j$.
The second term represents the intra-orbital Coulomb interaction with its magnitude $U$.
The third term represents the interorbital Coulomb interaction.
The fourth term represents Hund's coupling $J_\text{H}$ between spins $\bm{S}_{i,\gamma}=(S_{i,\gamma}^{x},S_{i,\gamma}^{y},S_{i,\gamma}^{z})$ at different orbitals with $S_{i,\gamma}^{a}=\frac{1}{2}\sum_{\tau,\tau'}c_{i,\gamma,\tau}^{\dag} \sigma_{\tau\tau'}^{a} c_{i,\gamma,\tau'}$.
The last term represents the on-site interorbital pair hopping with $P_{i,\gamma}=c_{i,\gamma,\uparrow} c_{i,\gamma,\downarrow}$.

In addition, we extend this Hamiltonian by introducing the following crystal-field (CF) splitting term
\begin{align}\label{eq-crysfield}
\mathcal{H}_\text{CF}=\frac{\Delta}{2} \sum_{i,\tau} (n_{i,a,\tau}-n_{i,b,\tau}).
\end{align}
Therefore, the total Hamiltonian $\mathcal{H}_\text{ETH}$ of an extended two-orbital Hubbard model (ETHM) is described by $\mathcal{H}_\text{ETH}=\mathcal{H}_\text{TH}+\mathcal{H}_\text{CF}$.
Following Refs.~\cite{Kunes2014b, Kunes2016}, we set the model parameters as $U=4$, $J_\text{H}=U/4$, and $(t_{aa},t_{bb})=(0.4,-0.2)$ in units of eV, which are used to capture the basic features of perovskite cobaltites.
Note that we introduce the on-site interorbital pair hopping term, which is ignored in Refs.~\cite{Kunes2014, Kunes2016, Geffroy2018, Nishida2019}.
At half filling, a band (Mott) insulator is stable for large (small) $\Delta$, while an excitonic insulator can be realized for intermediate $\Delta$.

The continuity equation along with the Heisenberg equation of motion for a spin operator $S^{z}_{l,\gamma}$ leads to a spin-current operator
\begin{align}
j_{\gamma \gamma'}^{s} (\bm{r}) := i \left(\text{sgn}~t_{\gamma \gamma'}\right) \sum_{\tau} \frac{s_\tau}{2} \left ( c_{l,\gamma,\tau}^{\dag} c_{m,\gamma',\tau} - c_{m,\gamma',\tau}^{\dag}c_{l,\gamma,\tau}  \right )
\label{eq:sc}
\end{align}
for a bond $(l,m)$ connecting sites $l$ and $m$ located at position vector $\bm{r}$, where $s_\tau=+1\,(-1)$ for $\tau=\uparrow\, (\downarrow)$.
To investigate the spin current, we use a pinning-field approach~\cite{Marston2002, Schollwock2003, Fjaerestad2006, White2007, Qin2020, Jiang2021}, where we introduce a small pinning field $j_{\gamma \gamma'}^{s} (\bm{r})$ on a bond $(l,m)$ of orbitals $\gamma$ and $\gamma'$ described by $\mathcal{H}^\text{s}_{\gamma \gamma'}=-h |t_{\gamma \gamma'}| j_{\gamma \gamma'}^{s} (\bm{r})$ with $h=0.0001$.
Note that conclusions obtained from this approach are essentially the same as those obtained from correlation functions~\cite{Schollwock2003, Fjaerestad2006}.
In Appendix~\ref{app-a}, we demonstrate that the pinning-field approach is useful for detecting numerically current correlations that appear in the staggered flux phase.
The pinning-field approach has an advantage because the DMRG method can calculate local quantities with much better accuracy than correlation functions~\cite{White2007, Jiang2021}.
Although the computational cost increases because complex numbers have to be used, the pinning field approach has been applied successfully for the DMRG study of off-diagonal orders or fluctuations of superconducting pairs and current, which are usually difficult to detect.

\begin{figure*}[tbh]
  \centering
    \includegraphics[clip, width=40pc]{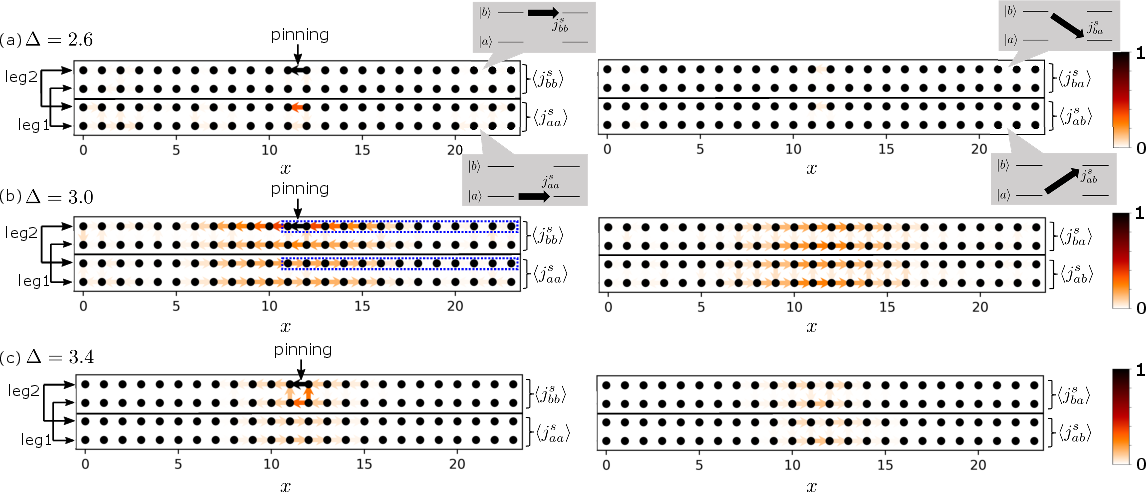}
    \caption{$\langle j^{s}_{\gamma \gamma'} (\bm{r})\rangle$ for the ETHM in the two-leg ladder with $(L_{x},L_{y})=(24,2)$ 
    at electron density $n=1.92$ close to half filling. 
    Their normalized magnitudes are shown by arrows with heatmap at the bond $\bm{r}$ 
    for the intra-orbital $(\gamma=\gamma')$ and interorbital $(\gamma \neq \gamma')$ spin current in the left and right panels, 
    respectively (also see schematic drawings). 
    Here, the legs of the ladder are labeled as leg~1 and leg~2, and the small pinning field is applied at the bond in leg 2 
    for orbital $b$ (indicated by ``pinning"). 
    The model parameters are set to $U=4$, $J_\text{H}=U/4$, $(t_{aa},t_{bb})=(0.4,-0.2)$, and $t_{ab}=t_{ba}=0.05$ with 
    (a) $\Delta=2.6$, (b) $\Delta=3$, and (c) $\Delta=3.4$ in units of eV. 
    The results indicated by blue dotted rectangles in (b) are also used in Fig.~\ref{Fig11}(a).
    }
    \label{Fig1}
\end{figure*}

Figure~\ref{Fig1} shows the results of $\langle j^{s}_{\gamma \gamma'}(\bm{r})\rangle$ for three different values of $\Delta=2.6$, 3, and 3.4, when the interorbital hoppings $t_{ab}=t_{ba}=0.05$ are introduced.
We choose $\Delta \simeq3$ since spontaneous hybridization between orbitals $a$ and $b$ is obtained (see Appendix~\ref{app-b}), indicating an exciton condensation.
Note that this type of interorbital hoppings satisfying $t_{ab}t_{ba}>0$ is usually referred to as ``even.''
These results are obtained near half filling, i.e., electron density $n=N/L=1.92$, where $N$ is the total number of electrons and $L=L_{x}L_{y}$ is the number of sites.
Note that $n=2$ corresponds to half filling in the ETHM.
We evaluate the ground state for the two-leg ladder with $(L_{x},L_{y})=(24,2)$, keeping $\chi=2500$ largest density-matrix eigenstates and taking 40 sweeps, which leads to a truncation error less than $10^{-11}$.
The pinning field is introduced in the bond of orbital $b$ that is indicated by an arrow with ``pinning'' in Fig.~\ref{Fig1}. 
Even when the pinning field is applied to orbital $a$, the following argument remains qualitatively the same.
Since the expectation values of the spin current depend on $h$ and increase in proportion to the density of states at the Fermi level, we normalize these quantities by $|\langle j^{s}_{\gamma \gamma'}(\bm{r})\rangle|$ at the bond applied with the pinning field in order to compare the results among different electronic states.

\begin{figure*}[tbh]
  \centering
    \includegraphics[clip, width=40pc]{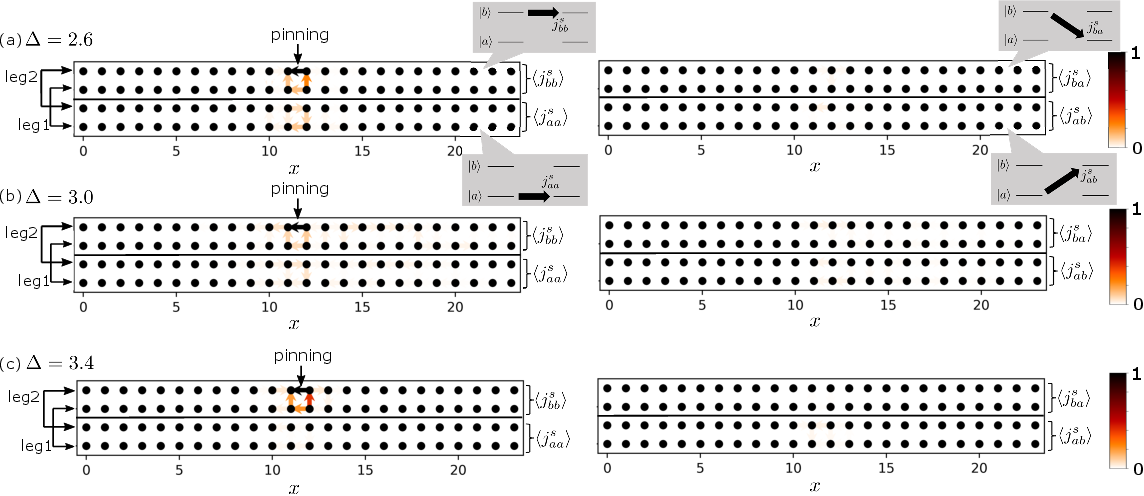}
    \caption{Same as Fig.~\ref{Fig1} but for $t_{ab}=-t_{ba}=0.05$.}
    \label{Fig2}
\end{figure*}

We find in Fig.~\ref{Fig1}(b) that the signal of sLC is most enhanced with clear sLC textures for $\Delta=3$, with which exciton condensation in the spin channel is associated~\cite{note1}.
The spatial distribution of $\langle j_{aa}^{s}(\bm{r})\rangle$ and $\langle j_{bb}^{s}(\bm{r})\rangle$ away from the bond with the pinning field, indicated by the blue dotted rectangles in Fig.~\ref{Fig1}(b), decays in distance and approximately follows the power-law behavior (see Appendix~\ref{app-c}).
For larger and smaller values of $\Delta$, i.e., $\Delta=2.6$ and $\Delta=3.4$, no sLC textures are observed in Figs.~\ref{Fig1}(a) and \ref{Fig1}(c).
Note that the presence of Hund's coupling is crucial for realizing the sLC texture, while the pair hopping is irrelevant. 
We also find that all off-diagonal parts of spin current $\langle j_{ab}^{s}(\bm{r})\rangle$ and $\langle j_{ba}^{s}(\bm{r})\rangle$, shown in Fig.~\ref{Fig1}(b), flow in the same direction.
At first glance, this behavior appears to be a spontaneous flow of global spin current.
However, it turns out that the total spin current including both orbital diagonal and off-diagonal parts vanishes~\cite{Geffroy2018, Nishida2019}, thus satisfying the Bloch theorem~\cite{Brillouin1933, Smith1935, Bohm1949, Watanabe2019}.

We can obtain sLC textures only when even-type interorbital hoppings are introduced.
In other words, no sLC textures emerge if odd-type interorbital hoppings are introduced or if no interorbital hoppings are introduced. 
Figure~\ref{Fig2} shows the results of $\langle j^{s}_{\gamma \gamma'}(\bm{r})\rangle$ for the odd-type interorbital hoppings $t_{ab}=-t_{ba}=0.05$. 
Indeed, we find that sLC textures are short-ranged.
We also obtain almost the same results as in Fig.~\ref{Fig2} when no interorbital hoppings are introduced.
Therefore, the symmetry of the interorbital hoppings is a key factor in the generation of sLC textures.
These features are in fact consistent with the previous work based on the (dynamical) mean-field theory~\cite{Kunes2016, Geffroy2018, Nishida2019}, and thus we conclude that a sLC texture associated with an exciton condensation in the spin channel can occur in the ETHM.

It should be noted, however, that our numerical results are consistent with previous studies only for the two-leg ladder system.
We find that the spatial distribution of the spin current away from the bond with the pinning field decays more steeply in the three- and four-leg ladder systems than in the two-leg ladder system.
Since an excitonic condensation in the spin channel has been proposed in Pr$_{0.5}$Ca$_{0.5}$CoO$_{3}$ and Ca$_{2}$RuO$_{4}$~\cite{Kunes2014b, Kunes2016}, there is a possibility of sLC textures being realized in these materials.
However, our DMRG study suggests that having a two-leg structure is also an important condition for the emergence of sLC textures.

\subsection{Carrier-doped orbital-selective Mott insulators}\label{sec3b}

\begin{figure*}[th]
  \centering
    \includegraphics[clip, width=40pc]{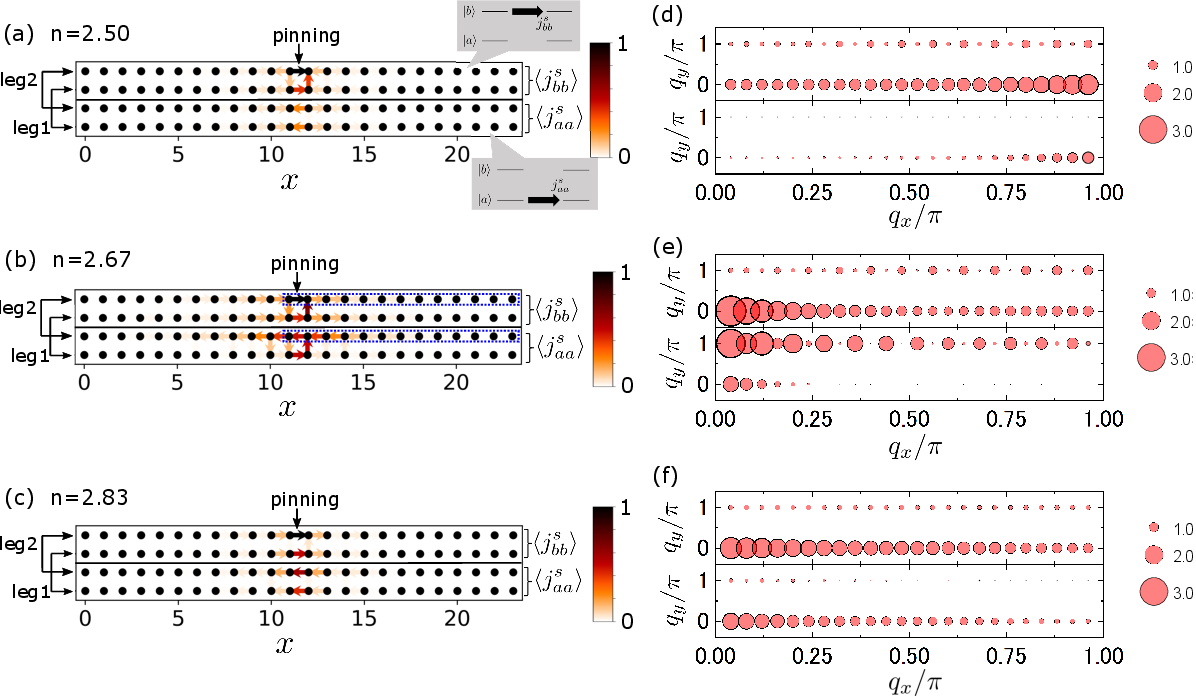}
    \caption{
    $\langle j^{s}_{\gamma \gamma'} (\bm{r})\rangle$ with $\gamma=\gamma'$ 
    for the ETHM in the two-leg ladder with $(L_{x},L_{y})=(24,2)$. 
    Their normalized magnitudes are shown by arrows with a heat map at the bond $\bm{r}$ for (a) $n=2.5$, (b) $n=2.67$, 
    and (c) $n=2.83$. Here, the legs of the ladder are labeled as leg 1 and leg 2, and the small pinning field is applied at the bond 
    in leg 2 for orbital $b$ (indicated by ``pinning"). The model parameters are set to $U=3.5$, 
    $J_\text{H}=U/4$, $(t_{aa},t_{bb})=(-0.5,-0.15)$, and $\Delta=-1.6$ in units of eV with the interorbital hoppings $t_{ab}=t_{ba}=0$. 
    The results indicated by blue dotted rectangles in (b) are also used in Fig.~\ref{Fig11}(a).
    In (d), (e), and (f), $J_{\gamma}(\bm{q})$ are evaluated from the results shown in (a), (b), and (c), respectively. 
    The diameter of the bubbles indicates the value of $J_{\gamma}(\bm{q})$ at different momentum $(q_x,q_y)$. 
    The lower and upper panels in each figure are for 
    $\langle j^{s}_{aa} (\bm{r}) \rangle$ and $\langle j^{s}_{bb} (\bm{r}) \rangle$ or 
    $J_a(\bm{q})$ and $J_b(\bm{q})$, respectively.
    }
    \label{Fig3}
\end{figure*}

In the previous section, we demonstrated that sLC textures emerge in the ETHM, and we showed that interorbital hoppings play a crucial role in stabilizing sLC textures associated with an exciton condensation in the spin channel.
In this section, we shall demonstrate that it is also possible to realize sLC textures in the ETHM without interorbital hoppings, i.e., $t_{ab}=t_{ba}=0$.
For this purpose, we set the model parameters to be $U=3.5$, $J_\text{H}=U/4$, $\Delta=-1.6$, and $(t_{aa},t_{bb})=(-0.5,-0.15)$ in units of eV, which are used to describe orbital-selective Mott insulators such as BaFe$_{2}$Se$_{3}$~\cite{Herbrych2019}.
As in Sec.~\ref{sec3a}, we consider the two-leg ladder with $(L_{x},L_{y})=(24,2)$.
Magnetic structures of this model have been investigated by the DMRG method and several types of block AFM order have been suggested~\cite{Herbrych2019}.

We calculate the ground state of the ETHM by using the DMRG method, keeping $\chi=2500$ largest density-matrix eigenstates and taking 40 sweeps, which leads to a truncation error less than $10^{-7}$.
Figures~\ref{Fig3}(a)--\ref{Fig3}(c) show the results of $\langle j^{s}_{aa}(\bm{r})\rangle$ and $\langle j^{s}_{bb}(\bm{r})\rangle$ for three different electron densities $n=2.5$, 2.67, and 2.83.
In these calculations, we introduce the pinning field at the bond of orbital $b$ indicated in the figure. 
However, the following argument remains qualitatively the same even when the pinning field is applied to orbital $a$. 
At $n=2.5$, a $(\pi,0)$ stripe order, i.e., AFM spin alignment along the legs and ferromagnetic (FM) spin alignment along the rungs, appears. 
Since the magnetic structure has already been studied~\cite{Herbrych2019}, here we focus on the possibility of the emergence of sLC textures.

First, we do not find robust sLC textures for typical values of electron density in the range of $2.0\lesssim n \lesssim 2.5$ and $n\sim 3$.
Indeed, the correlation of spin current is short-ranged, as shown in Fig.~\ref{Fig3}(a) for $n=2.5$ and Fig.~\ref{Fig3}(c) for $n=2.83$. 
These features are better quantified by evaluating the Fourier transform of these quantities, i.e., $J_{\gamma}(\bm{q})=\sum_{\bm{r}}\langle j^{s}_{\gamma \gamma}(\bm{r})\rangle \cos(\bm{q}\cdot \bm{r})$.
As shown in Fig.~\ref{Fig3}(d) for $n=2.5$ and Fig.~\ref{Fig3}(f) for $n=2.83$, $J_{\gamma}(\bm{q})$ has a structure around $\bm{q}=(\pi,0)$ and $(0,0)$, respectively, but it is rather broad.
In contrast, we find the enhanced signal of sLC textures in the range of electron density $2.63<n<2.71$, where the correlation of spin current is longer-ranged, as shown in Figs.~\ref{Fig3}(b) and \ref{Fig3}(e) for $n=2.67$.
The spatial distribution of $\langle j_{aa}^{s}(\bm{r})\rangle$ and $\langle j_{bb}^{s}(\bm{r})\rangle$ away from the bond with the pinning field, indicated by the blue dotted rectangles in Fig.~\ref{Fig3}(b), decays in distance and approximately follows the power-law behavior (see Appendix~\ref{app-c}).
It should be noted that the correlation of the spin current in Fig.~\ref{Fig3}(b) appears weaker than that in Fig.~\ref{Fig1}(b).

We should note that the sLC textures found here are unaffected by the introduction of the interorbital hoppings $t_{ab}=\pm t_{ba}=0.05$ also used in Sec.~\ref{sec3a}, which indicates that the sLC textures found in this section are due to a mechanism different from the exciton condensation in the spin channel. 
We also find that the introduction of $\Delta \neq 0$ does not play an essential role.
Since the sLC signal becomes small when the ratio of $t_{bb}/t_{aa}$ is closer to 1, the difference in the itinerancy of electrons in orbitals $a$ and $b$ is important to the development of sLC textures.

Recalling that noncollinear magnetism can produce spin current~\cite{Katsura2005}, vector chirality is one of the possible origins for the sLC textures.
It has been proposed in Ref.~\cite{Sroda2021} that the correlation of vector chirality $j^\text{ns}_{bb}(\bm{r}):= \left( \bm{S}_{l,b} \times \bm{S}_{m,b}\right)_{z} = S^{x}_{l,b}S^{y}_{m,b}-S^{y}_{l,b}S^{x}_{m,b}$ of localized spins on orbital $b$ can be developed in the GKHM, which is an effective model of the ETHM in the strong-coupling limit.
Note that the vector chirality operator $j^\text{ns}_{bb}(\bm{r})$ is also regarded as a spin-current operator defined on localized spins.
It is also interesting to notice that the electron density $n$ where sLC textures emerge is consistent for both models, i.e., the ETHM studied in this paper and the GKHM studied in Ref.~\cite{Sroda2021}. 
However, the pattern of sLC textures is different between these two systems. 
In the ETHM, spin current flows in the leg direction for each orbital $a$ or $b$, as shown in Fig.~\ref{Fig3}(b).
Because spin currents in these two orbitals flow in opposite directions, the global spin current does not flow in total. 
In the GKHM, on the other hand, spin currents generated by itinerant electrons and by localized spins are inequivalent, which may thus induce another kind of sLC textures, i.e., staggered spin current circulating around $2\times2$ plaquettes~\cite{Sroda2021}.
We have also examined $j^\text{ns}_{\gamma \gamma}(\bm{r})$ in the ETHM and confirmed that the correlation of $j^\text{ns}_{\gamma \gamma}(\bm{r})$ is developed when $j^\text{s}_{\gamma \gamma}(\bm{r})$ exhibits enhanced correlation.

\begin{figure}[t]
  \centering
    \includegraphics[clip, width=20pc]{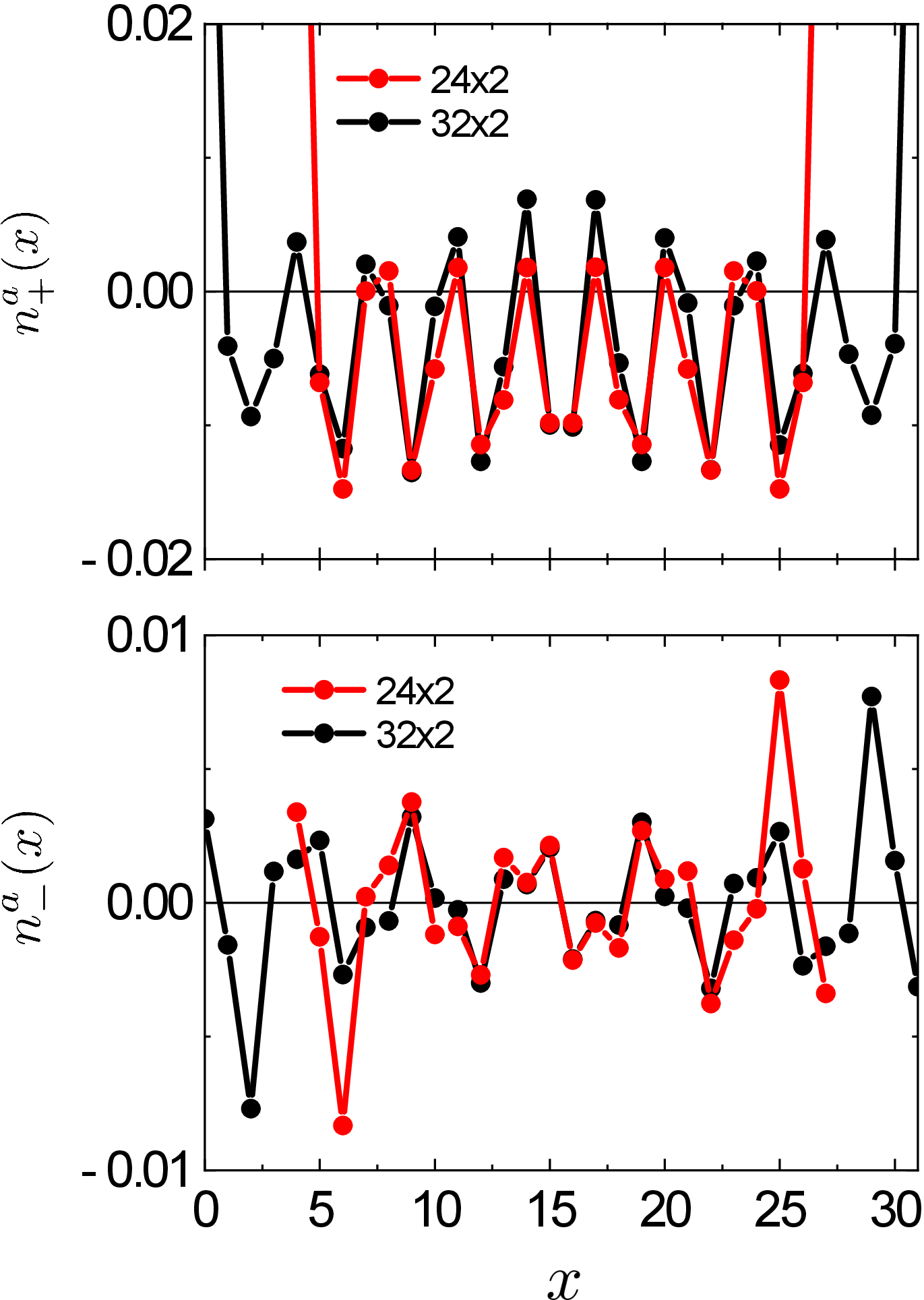}
    \caption{$n^{a}_{+}(x)$ (upper panel) and $n^{a}_{-}(x)$ (lower panel) for the ETHM in the two-leg ladder.
    Red circles are for $(L_x,L_y)=(24,2)$ at $n=8/3\simeq2.67$ and black circles are for $(L_x,L_y)=(32,2)$ at 
    $n=85/32\simeq2.66$. The model parameters are the same as those used in Fig.~\ref{Fig3}
    }
    \label{Fig4}
\end{figure}

Finally, we note that the sLC textures found here in the ETHM coexist with charge stripes.
Figure~\ref{Fig4} shows the results of $n^{a}_{+}(x) := \frac{1}{L_{y}}\sum_{y=1,2}\left( \langle n_{x,a}^{y}\rangle-n\right) $ and $n^{a}_{-}(x) := \frac{1}{L_{y}}\sum_{y=1,2}\left[  (-1)^{y}\langle n_{x,a}^{y}\rangle \right] $, where $n_{x,a}^{y}$ is an electron density operator at the $x$th rung ($x=1,2,\dots,L_x$) for leg $y$ and orbital $a$. 
These two quantities $n^{a}_{+}(x)$ and $n^{a}_{-}(x)$ represent, respectively, the average and difference of the numbers of electrons at legs 1 and 2 in each rung.
When the sLC textures emerge, we find that charge stripes also appear probed in both $n^{a}_{+}(x)$ and $n^{a}_{-}(x)$, as shown in Fig.~\ref{Fig4} for $n=2.67$ with $(L_x,L_y)=(24,2)$ (red circles) and $n=2.66$ with $(L_x,L_y)=(32,2)$ (black circles), indicating the spontaneous formation of charge stripes both along rungs and legs. 
After removing the contributions from edges to reduce the finite-size effects, we find that $\tilde{n}^{a}_{+}(q_{x}):=\sum_{x}n^{a}_{+}(x)\cos (q_{x}x)$ shows a peak structure at $q_{x}\sim 2$ and similarly $\tilde{n}^{a}_{-}(q_{x}):=\sum_{x}n^{a}_{-}(x)\cos (q_{x}x)$ shows a peak structure at $q_{x}\sim 1$, which correspond to charge stripes with the period of $\lambda \simeq 3$ and 6 (in units of the lattice constant), respectively. 
These charge stripes can trigger vector chirality when they form superlattice structures that break local inversion symmetry~\cite{Zhang2014, Fischer2023}.
Indeed, the emergence of antisymmetric exchange, i.e., the Dzyaloshinskii-Moriya interaction, has been proposed in ABC-type superlattices~\cite{Ham2021}.
The coexistence of vector chirality and charge stripes is one manifestation of multiferroicity~\cite{Katsura2005, Cheong2007}. 
We should also note that the coexistence of cLC textures and stripes has recently been reported in a spinless Hubbard model~\cite{JuliaFarre2023}.

\section{Single-orbital Hubbard model on a square lattice}\label{sec4}

In Sec.~\ref{sec3}, we have demonstrated the emergence of sLC textures in the two-orbital Hubbard model on a ladder lattice. 
Even with only a single orbital, the two-dimensional Hubbard model exhibits very rich quantum phases with highly entangled spin and charge degrees of freedom.
In this section, we focus on sLC textures in the single-orbital Hubbard model on a square lattice.
There are several proposals for sLC textures in the Hubbard model on a square lattice.
In the Hubbard model with a single hole on a square lattice, the emergence of sLC textures has been suggested~\cite{Zheng2018}.
The sLC textures are driven by a many-body Berry-like phase, i.e., phase string~\cite{Sheng1996, Weng1997, Shinjo2021a} in the single-hole $t$-$J$ model on a square lattice.
We should, however, note that total $S^{z}$ is nonzero in Ref.~\cite{Zheng2018}, where time-reversal symmetry is explicitly broken in the Hamiltonian.
Even for total $S^{z}$ being zero, the recent theoretical analysis based on the functional renormalization-group method in Ref.~\cite{Kontani2021} has revealed the emergence of sLC textures in a hole-doped Hubbard model on a square lattice with further-neighbor hoppings.  
The sLC textures found in this analysis are characterized with a wave vector $\bm{q}=(\pi/2,\pi/2)$, which is diagonal and closely related to the nesting vector of the Fermi surface~\cite{Kontani2021}.

\begin{figure*}[th]
  \centering
    \includegraphics[clip, width=40pc]{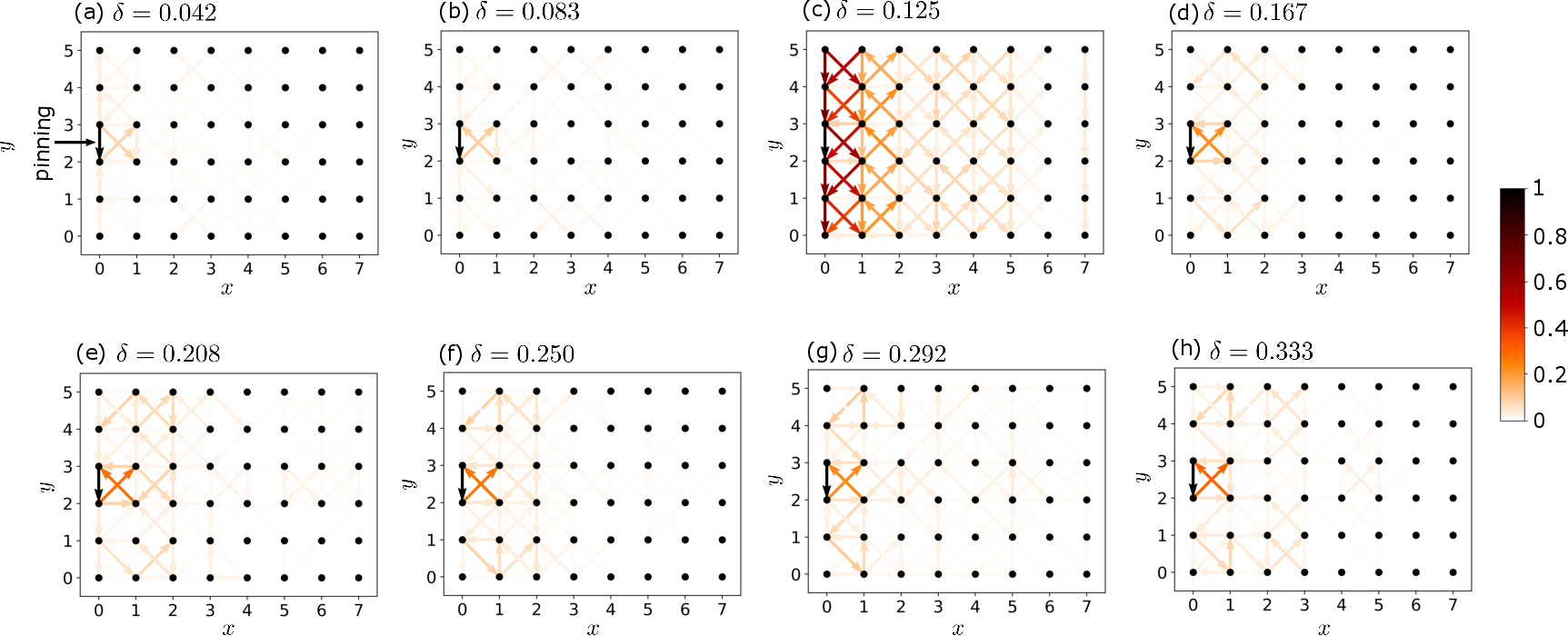}
    \caption{
    $\langle j^{s} (\bm{r})\rangle$ for the single-orbital Hubbard model with $U/t=9$ and $t'/t=-0.3$ 
    on the square lattice with $(L_{x},L_{y})=(8,6)$. 
    Their normalized amplitudes are shown by arrows with a heat map at the bond $\bm{r}$ for (a) $\delta=0.042$, (b) $\delta=0.083$, 
    (c) $\delta=0.125$, (d) $\delta=0.167$, (e) $\delta=0.208$, (f) $\delta=0.25$, (g) $\delta=0.292$, and (h) $\delta=0.333$. 
    Here, the bond to which the small pinning field is applied is indicated by ``pinning" in (a). 
    The same pinning field is also applied in (b)--(h), but it is not explicitly indicated.
    } 
    \label{Fig5}
\end{figure*}

Here, we investigate sLC textures in the single-orbital Hubbard model on a square lattice by using the DMRG method.
The Hamiltonian of the Hubbard model on a square lattice is given as 
\begin{align}
\mathcal{H}_{t\text{-}t'\text{-}U}=&-t\sum_{\langle i,j\rangle,\tau} (c_{i,\tau}^{\dag}c_{j,\tau} + \text{H.c.}) \nonumber \\
&-t'\sum_{\langle \langle i,j\rangle \rangle,\tau} (c_{i,\tau}^{\dag}c_{j,\tau} + \text{H.c.})
+U\sum_{i}n_{i,\uparrow}n_{i,\downarrow},
\end{align}
where $c_{i,\tau}^\dag$ is the electron creation operator at site $i$ with spin $\tau\,(=\uparrow,\downarrow)$ and $n_{i,\tau}=c_{i,\tau}^\dag c_{i,\tau}$.   
$t$ and $t'$ are the nearest-neighbor and next-nearest-neighbor hoppings on a square lattice, respectively, and $U$ is the on-site Coulomb interaction. 
The sum indicated by $\langle i,j  \rangle$ ($\langle \langle i,j \rangle \rangle$) runs over all pairs of nearest-neighbor (next-nearest-neighbor) sites $i$ and $j$.

We use a cluster of $(L_{x},L_{y})=(8,6)$ on a cylinder geometry, i.e., open and periodic boundary conditions along $x$ and $y$ axes, respectively. 
Although we can treat even larger clusters at the expense of accuracy, we avoid using too large clusters since the high computational accuracy is required to correctly calculate off-diagonal quantities such as spin current. 
To treat a two-dimensional cluster in the DMRG method, we construct a snakelike one-dimensional chain out of the two-dimensional square lattice, running from site at $(0,0)$ to site at $(0,L_{y}-1)$, then from site at $(1,L_{y}-1)$ to site at $(1,0)$, and this pattern is repeated until we reach site at $(L_{x}-1,0)$.
We keep $\chi=10000$ largest density-matrix eigenstates and take 40 sweeps in the DMRG calculations, leading to a truncation error less than $5\times 10^{-5}$.

Similar to Eq.~(\ref{eq:sc}), the spin current operator for the single-band Hubbard model is defined as 
\begin{align}
j^{s} (\bm{r}) := i \left(\text{sgn}~t\, {\rm or}\,t'\right) \sum_{\tau} \frac{s_\tau}{2} \left ( c_{l,\tau}^{\dag} c_{m,\tau} - c_{m,\tau}^{\dag}c_{l,\tau}  \right )
\label{eq:sc2}
\end{align}
for a bond $(l,m)$ connecting sites $l$ and $m$ located at a position vector $\bm{r}$. 
To investigate the spin current, we introduce a small pinning field $j^{s} (\bm{r})$ on a bond $(l,m)$, described by $\mathcal{H}^\text{s}=-h |t| j^{s} (\bm{r})$ with $h=0.0001$, i.e., site $l$ located at $(0,2)$ and site $m$ located at $(0,3)$ for our cluster with $(L_x,L_y)=(8,6)$ (also see Fig.~\ref{Fig5}).

Figure~\ref{Fig5} summarizes the results of $\langle j^{s}(\bm{r}) \rangle$ for $t'/t=-0.3$ and $U/t=9$ with different hole concentrations $\delta=1-n$.
We find that the correlation of $\langle j^{s}(\bm{r}) \rangle$ is rather short-ranged except for $\delta=0.125$. 
Note that half filling is achieved at $n=1$ for the single-orbital Hubbard model.
We also note that the global spin current in the $x$ direction should be zero due to the open boundary conditions, whereas the spin current in the $y$ direction should be suppressed by $L_{y}^{-1}$ due to the periodic boundary conditions~\cite{Watanabe2019}.

\begin{figure*}[t]
  \centering
    \includegraphics[clip, width=40pc]{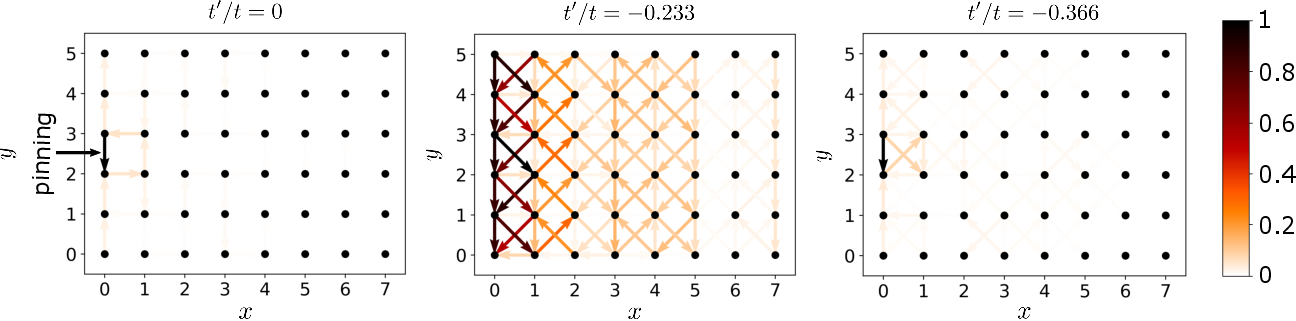}
    \caption{
    Same as Fig.~\ref{Fig5} but for three different $t'$ values (indicated in the figures) at $\delta=0.125$
    }
    \label{Fig6}
\end{figure*}

\begin{figure*}[t]
  \centering
    \includegraphics[clip, width=40pc]{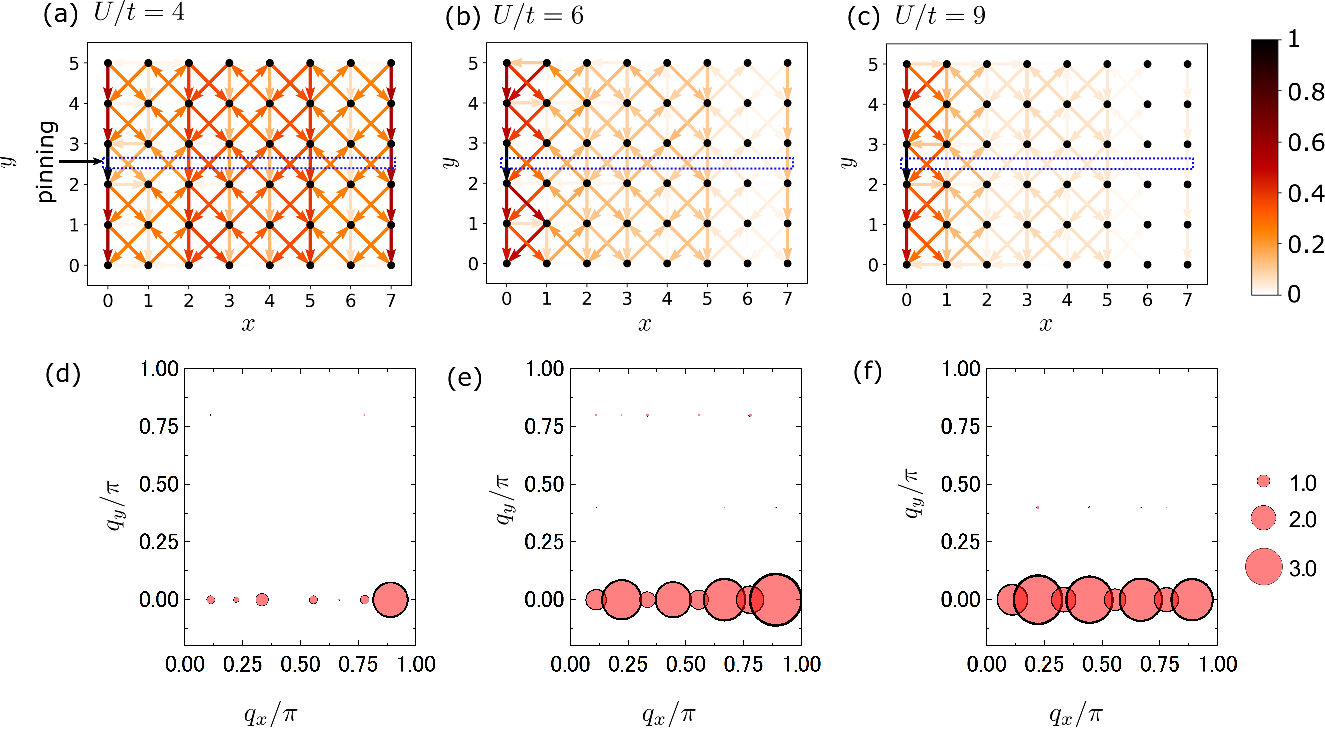}
    \caption{
    (a)--(c) Same as Fig.~\ref{Fig5} but for (a) $U/t=4$, (b) $U/t=6$, and (c) $U/t=9$ at $\delta=0.125$ and $t'/t=-0.266$. 
    The results indicated by blue dotted rectangles are also used in Fig.~\ref{Fig11}(b).
    (d)--(f) $J(\bm{q})$ evaluated from $\langle j^{s}(\bm{r}) \rangle$ shown in (a),(b), i.e., 
    for (d) $U/t=4$, (e) $U/t=6$, and (f) $U/t=9$. The diameters of bubbles indicate 
    the values of $J(\bm{q})$.
    }
    \label{Fig7}
\end{figure*}

In addition to the hole concentration, several other conditions are required for the emergence of the sLC textures.
Figure~\ref{Fig6} shows the results of $\langle j^{s}(\bm{r}) \rangle$ for different values of $t'$ at $\delta=0.125$, revealing that the presence of $t'/t\sim -0.25$ is necessary to induce the robust sLC textures.
Figures~\ref{Fig7}(a)--\ref{Fig7}(c) show the results of $\langle j^{s}(\bm{r}) \rangle$ for three different values of $U/t=4, 6$, and 9 at $\delta=0.125$ and $t'/t=-0.266$.
These results clearly find that the sLC textures are most extended and enhanced when $U/t$ is smaller. 
Figures~\ref{Fig7}(d)--\ref{Fig7}(f) show $J(\bm{q})=\sum_{\bm{r}}\langle j^{s}(\bm{r})\rangle \cos(\bm{q}\cdot \bm{r})$ evaluated from the results of $\langle j^{s}(\bm{r}) \rangle$ in Figs.~\ref{Fig7}(a)--\ref{Fig7}(c).
We find that the centroid of $J(\bm{q})$ is concentrated toward $\bm{q}\simeq(\pi,0)$ with decreasing $U/t$. 
We should also note that even when $U/t=4$, no sLC textures emerge if $t'/t$ deviates significantly away from $-0.25$.

It is interesting to compare the sLC textures obtained here by the DMRG method with those reported in Ref.~\cite{Kontani2021}.
Based on the functional-renormalization group method, the presence of sLC textures has been proposed in the single-orbital Hubbard model with $t'/t=-1/6$, $t''/t=1/5$, and $U/t=3.3$ at $\delta=0.2$, where $t''$ is the third-nearest-neighbor hopping~\cite{Kontani2021}. 
The suggested sLC textures are characterized by wave vector $\bm{q}\simeq(\pi/2,\pi/2)$.
The similarity and differences of the results between our study and their study~\cite{Kontani2021} are summarized as follows. 
Both studies suggest that the introduction of relatively small or intermediate $U/t$ and $t'/t$ is crucial for the emergence of the robust sLC textures. 
However, the sLC textures appear most significantly at $\delta=0.125$ in our study but at $\delta=0.2$ in their study. 
The wave vector $\bm{q}$ characterizing the spatial pattern of sLC textures is also different: while the axial-sLC textures with $\bm{q}=(\pi,0)$ are found by our DMRG study, the diagonal-sLC textures with $(\pi/2,\pi/2)$ are obtained by their functional renormalization group study~\cite{Kontani2021}.
The axial-sLC textures may be stabilized here because we employ the cluster of a cylinder geometry. 
However, a more detailed study is warranted to clarify this point.

Now we comment on the coexistence of sLC textures and charge stripes at $\delta=0.125$.
The hole density of $\delta=0.125$ is well-known as the density at which the charge stripes appear, and many previous studies have been focused on this density~\cite{Hu2012, Marino2022, White1998, White2000, White2004, White2009, Jiang2019, Corboz2011, Corboz2014, Dong2020, Zheng2017, Tohyama2018, Tohyama2020}.
In the Hubbard model in the strong-coupling region or the $t$-$J$ model on a four-leg ladder under the cylinder geometry, the axial charge stripes with a period of $\lambda=8$ (in units of the lattice constant), characterized by its ordering wave vector $\bm{q}=(\pi/4,0)$, are stabilized at $\delta=0.125$~\cite{Jiang2019}.
Introducing $t'/t=-0.25$, the $\lambda=4$ charge stripes with their ordering wave vector $\bm{q}=(\pi/2,0)$ appear. 
Here, we shall show that the charge stripes also appear on the six-leg ladder used in this section.

\begin{figure}[t]
  \centering
    \includegraphics[clip, width=20pc]{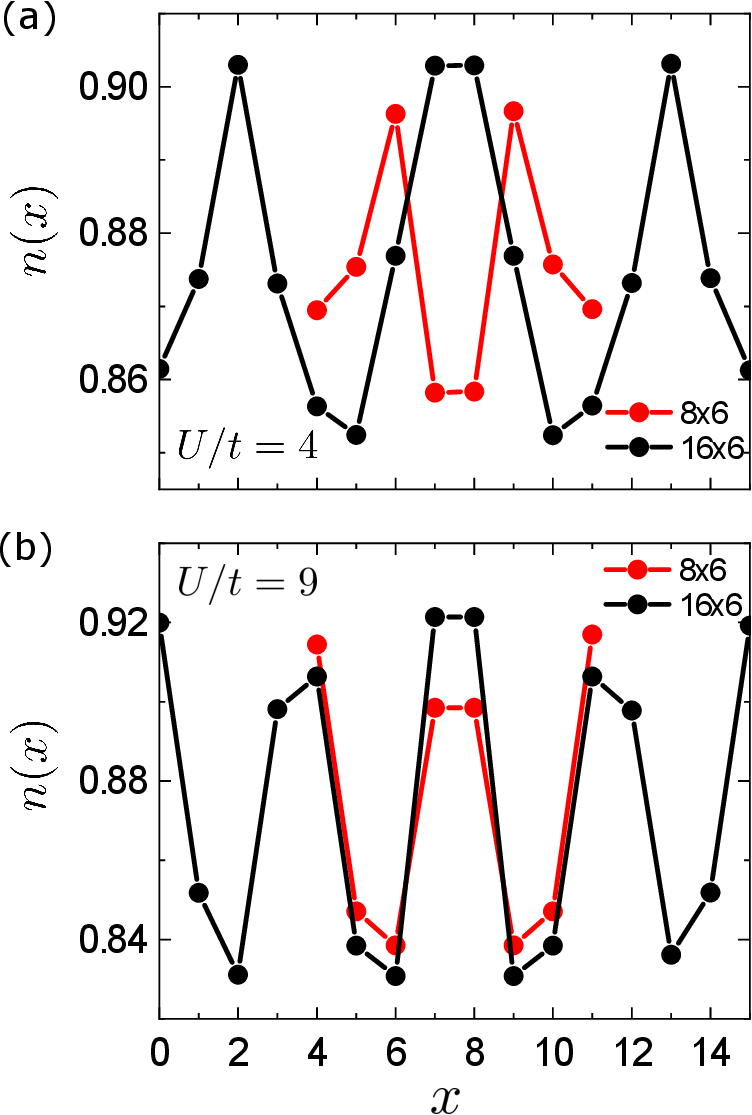}
    \caption{
    Charge density averaged over sites along the $y$ direction, $n(x)$, for the single-orbital Hubbard model with $t'/t=-0.266$ 
    at $\delta=0.125$ on the square lattices with $(L_{x},L_{y})=(8,6)$ (red circles) and $(L_{x},L_{y})=(16,6)$ (black circles). 
    The on-site Coulomb interaction is set to (a) $U/t=4$ and (b) $U/t=9$. 
    For easier comparison, the results for $(L_{x},L_{y})=(8,6)$ are displaced by 4 in the horizontal axis.
    }
    \label{Fig8}
\end{figure}

To quantify the charge distribution along the $x$ axis, i.e., the leg direction, we evaluate $n(x):=\frac{1}{L_{y}}\sum_{y=0}^{L_y-1}\langle n_{x}^y\rangle$, where $n_{x}^{y}$ is an electron density operator at the $x$th rung ($x=0,1,\dots,L_x-1$) in leg $y$.
Since the computation of diagonal quantities such as the charge density is not severely sensitive to the lower accuracy in the DMRG method, we can evaluate this quantity $n(x)$ for a larger cluster with $(L_{x},L_{y})=(16,6)$. 
Since the cluster is on a cylinder geometry, we expect that the axial charge stripes are more stable. 
We indeed find in Fig.~\ref{Fig8} that the axial charge stripes appear at $\delta=0.125$, and the period of these stripes becomes longer with decreasing $U$.

\begin{figure}[t]
  \centering
    \includegraphics[clip, width=20pc]{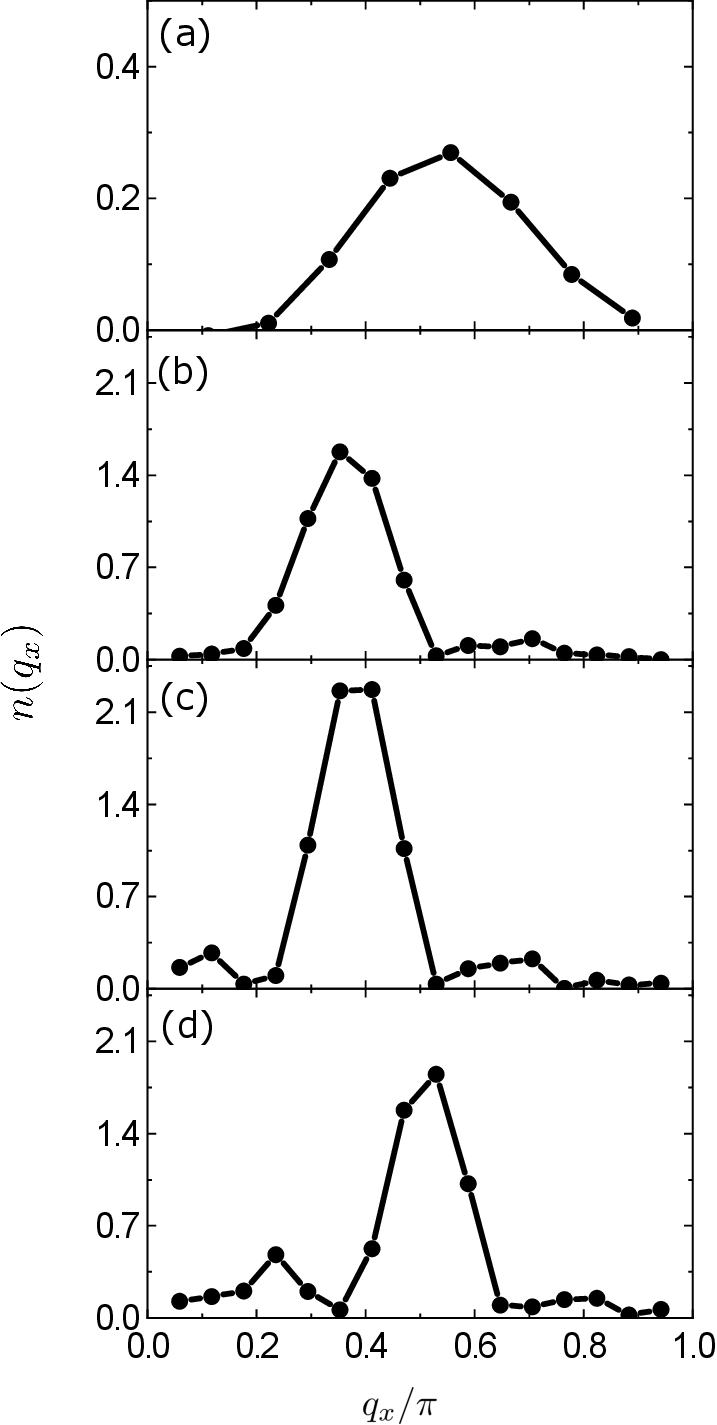}
    \caption{
    $n(q_{x})$ for the single-orbital Hubbard model with $t'/t=-0.266$ at $\delta=0.125$.
    (a) $U/t=4$ on the square lattice with $(L_{x},L_{y})=(8,6)$, and (b) $U/t=4$, (c) $U/t=6$, and (d) $U/t=9$ on the 
    square lattice with $(L_{x},L_{y})=(16,6)$.
    }
    \label{Fig9}
\end{figure}

To further discuss the period of the charge stripes, we also evaluate $n(q_{x}):= \sum_{x} n(x) \cos (q_{x} x)$ and the results are shown in Fig.~\ref{Fig9}.
As shown in Fig.~\ref{Fig9}(a), when $t'/t= -0.3$ and $U/t=4$, $n(q_{x})$ has a broad peak at $0.32\lesssim q_{x}/\pi \lesssim 0.80$ for the cluster with $(L_{x},L_{y})=(8,6)$.
We can reduce the finite-size effect when we consider the cluster with $(L_{x},L_{y})=(16,6)$, for which the results of $n(q_{x})$ are shown in Figs.~\ref{Fig9}(b)--\ref{Fig9}(d) for three different values of $U/t=4$, 6, and 9.
As shown in Fig.~\ref{Fig9}(b), when $t'/t= -0.3$ and $U/t=4$, $n(q_{x})$ exhibits a peak at $0.25 \lesssim q_{x}/\pi \lesssim 0.48$, which indicates the charge stripes with $\lambda \simeq5$.
This is consistent with the results obtained by the variational Monte-Carlo study on a six-leg Hubbard ladder with $t'$~\cite{Marino2022}. 
With increasing $U$, the characteristic ordering wave vector of the charge stripes becomes larger, as shown in Fig.~\ref{Fig9}(d) for $U/t=9$, where $n(q_{x})$ has a peak at $0.45 \lesssim q_{x}/\pi \lesssim 0.60$, leading to $\lambda \simeq4$.
Since $\delta=0.125$ is a key factor in the development of the sLC textures, it is most likely that the presence of $\lambda \simeq4$ and 5 charge stripes is crucial for the emergence of the sLC textures found here. 
Assuming that a symmetry breaking leads to the emergent spin-orbit coupling as discussed in Sec.~\ref{sec2}, the electric field produced locally by charge stripes may induce spin current by a similar mechanism to the spin Hall effect~\cite{Xiao2010, Sinova2015}.

\section{Summary}\label{sec5}

We have studied sLC textures emerging in the ground states of the Hubbard models by using the DMRG method. 
In Particular, we have investigated carrier-doped (i) excitonic insulators, (ii) orbital-selective Mott insulators, and (iii) two-dimensional Mott insulators, modeled by the ETHM on a two-leg ladder lattice in (i) and (ii), and the single-orbital Hubbard model with the next-nearest hopping 
$t'$ on a square lattice in (iii).  
In these systems, we have obtained the enhanced sLC textures developed around a bond to which the pinning field is applied. 

In system (i), we have found the emergence of sLC textures which is associated with an exciton condensation in the spin channel. 
Using model parameters motivated for excitonic insulators, we have found that the sLC correlations are developed most significantly near half-filling at electron density $n=1.92$ when the crystal field and the interorbital hoppings are suitably introduced.
In system (ii), we have used typical model parameters for iron oxides such as BaFe$_{2}$Se$_{3}$ and found that the robust sLC textures emerge in the ETHM without introducing any interorbital hoppings.
The sLC textures are developed most profoundly when a relatively large number of carriers is introduced in the range of electron density $2.63<n<2.71$ and the difference in itinerancy of electrons in the two orbitals is large.
We have also found that the sLC textures coexist with the charge stripes formed in both rungs and legs. 
In system (iii), we have found that the sLC textures are most enhanced and extended at $\delta=0.125$ when $t'/t \sim -0.25$ is introduced. 
We also found that the sLC textures are most developed when $U$ decreases from $U/t=9$ to $U/t=4$.
The $\lambda \simeq4$ and 5 charge stripes also simultaneously appear when the sLC textures emerge. 

In conclusion, we have found the conditions under which the sLC textures are developed in each of the three systems (i)--(iii). 
Our results clearly demonstrate that quantum many-body effects can induce local spin current in the ground state.
It is interesting to explore whether the sLC textures can lead to the development of spintronics in strongly correlated electron systems.
Finally, we note that the ground states with only short-ranged cLC correlations have been found in all three systems (i)--(iii).
Namely, the expectation value of the charge current away from the bond with the pinning field is nearly zero.


\begin{acknowledgments}

This work was supported by Grants-in-Aid for Scientific Research (B) (No.~19H01829, No.~19H05825, No.~20H01849, and No.~21H03455) and a Grant-in-Aid for Early-Career Scientists (No.~23K13066) from Ministry of Education, Culture, Sports, Science, and Technology (MEXT), Japan, and by JST PRESTO (Grant No. JPMJPR2013).
This work was also supported in part by the COE research grant in computational science from Hyogo Prefecture and Kobe City through Foundation for Computational Science. 
Numerical calculation was carried out using the HOKUSAI supercomputer at RIKEN, the facilities of the Supercomputer Center at Institute for Solid State Physics, the University of Tokyo, and the supercomputer system at the information initiative center, Hokkaido University through the HPCI System Research Project (Project ID: hp220049).

\end{acknowledgments}

\appendix

\section{Pinning-field approach}\label{app-a}

In Secs.~\ref{sec3} and ~\ref{sec4}, we introduce a small pinning field $h$ to the systems to study the spatial distribution of sLC correlations. 
The pinning field approach has been used previously to investigate cLC correlations~\cite{Schollwock2003, Fjaerestad2006}, and in this appendix we shall reproduce their results of the cLC texture in our DMRG calculations to clarify the role of the pinning field. 

The Hamiltonian exhibiting a cLC texture reads 
\begin{align}\label{eq-appA}
\mathcal{H}_{t\text{-}J\text{-}V}=&-t\sum_{\langle i,j\rangle,\tau} (\tilde c_{i,\tau}^{\dag} \tilde c_{j,\tau}+\text{H.c.}) \nonumber \\
&+ J\sum_{\langle i,j\rangle} \left( \tilde{\bm{S}}_{i}\cdot \tilde{\bm{S}}_{j} -\frac{1}{4}\tilde n_{i} \tilde n_{j} \right)\nonumber \\
&+V_{1}\sum_{\langle i,j\rangle}\tilde n_{i}\tilde n_{j} + V_{2}\sum_{\langle \langle i,j\rangle \rangle} \tilde n_{i} \tilde n_{j},
\end{align}
which is a $t$-$J$ model with the (next)-nearest-neighbor interaction $V_{1}$ ($V_{2}$) on a two-leg ladder lattice with open boundary conditions.
Here, $\tilde c_{i,\tau} = c_{i,\tau}(1-n_{i,-\tau})$, $c_{i,\tau}$ is the annihilation operator of an electron with spin $\tau\,(\uparrow,\downarrow)$ at site $i$, and $n_{i,\tau}=c_{i,\tau}^\dag c_{i,\tau}$ with $-\tau$ being the opposite spin of $\tau$. 
$\tilde n_{i}=\sum_{\tau}\tilde n_{i,\tau}$ with $\tilde n_{i,\tau}=\tilde c_{i,\tau}^{\dag} \tilde c_{i,\tau}=n_{i,\tau}(1-n_{i,-\tau})$, and $(\tilde{\bm{S}}_{i})_a=\frac{1}{2}\sum_{\tau,\tau'}\tilde c_{i,\tau}^\dag\sigma_{\tau\tau'}^a \tilde c_{i,\tau'}=\frac{1}{2}\sum_{\tau,\tau'} c_{i,\tau}^\dag\sigma_{\tau\tau'}^a c_{i,\tau'}$ is the $a\,(=x,y,z)$ component of the spin operator at site $i$.
A charge-current operator for a bond $(l,m)$ connecting sites $l$ and $m$ located at a position vector $\bm{r}$ is defined as
\begin{align}
j^{c} (\bm{r}) := i \left(\text{sgn}~t\right) \sum_{\tau} \left ( \tilde c_{l,\tau}^{\dag} \tilde c_{m,\tau} - \tilde c_{m,\tau}^{\dag} \tilde c_{l,\tau}  \right ).
\label{eq:cc}
\end{align}

To investigate charge current, we introduce a small pinning field described by $\mathcal{H}^\text{c}=-h |t| j^{c} (\bm{r})$ with $h=0.0001$. 
Notice that the pinning field is applied only at the single bond $\bm{r}$.
The results for the spatial distribution of $\langle j^{c} (\bm{r}) \rangle$ on a two-leg ladder lattice with $(L_x,L_y)=(20,2)$ are summarized in Fig.~\ref{Fig10}, where we set $J/t=0.4$, $V_{1}/t=3$, and $V_{2}/t=1$. 
The bond with the pinning field is indicated by ``pinning'' in Fig.~\ref{Fig10}.
We find that indeed a cLC texture emerges at hole density $\delta=0.1$, exhibiting staggered flow of charge current, which is in good accordance with the staggered-flux order reported in Ref.~\cite{Schollwock2003}.
The introduction of finite $V_{1}$ and $V_{2}$ is a key ingredient to induce the cLC texture in this system.

\begin{figure}[t]
  \centering
    \includegraphics[clip, width=20pc]{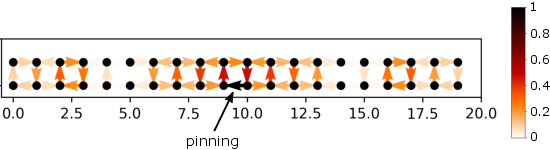}
    \caption{
    $\langle j^{c} (\bm{r})\rangle$ for the $t$-$J$ model $\mathcal{H}_{t\text{-}J\text{-}V}$ given in Eq.~(\ref{eq-appA}) with $J/t=0.4$, $V_{1}/t=3$, and $V_{2}/t=1$ 
    on the two-leg ladder with $(L_{x},L_{y})=(20,2)$ at hole density $\delta=0.1$.
    Their normalized amplitudes are shown by arrows with a heat map at the bond $\bm{r}$. 
    The bond to which the small pinning field is applied is indicated by ``pinning."
    }
    \label{Fig10}
\end{figure}

\section{Hybridization induced by exciton condensation}\label{app-b}

\begin{figure}[t]
  \centering
    \includegraphics[clip, width=20pc]{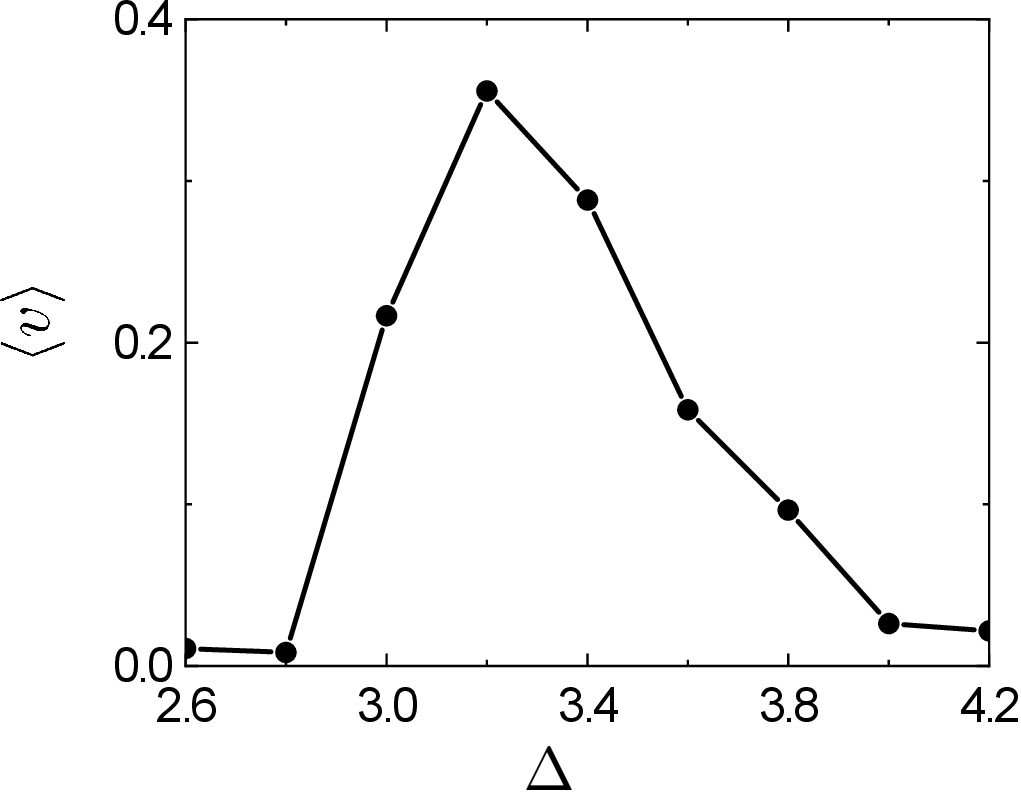}
    \caption{
    Hybridization average $\langle v \rangle$ between orbitals $a$ and $b$ as a function of the crystal-field splitting $\Delta$ for the ETHM in the two-leg ladder with $(L_{x},L_{y})=(24,2)$ at electron density $n=1.92$ close to half filling.
    The model parameters are set to $U=4$, $J_\text{H}=U/4$, $(t_{aa},t_{bb})=(0.4,-0.2)$, and $t_{ab}=t_{ba}=0.05$ in units of eV.
    }
    \label{Fig10}
\end{figure}

\begin{figure}[th]
  \centering
    \includegraphics[clip, width=20pc]{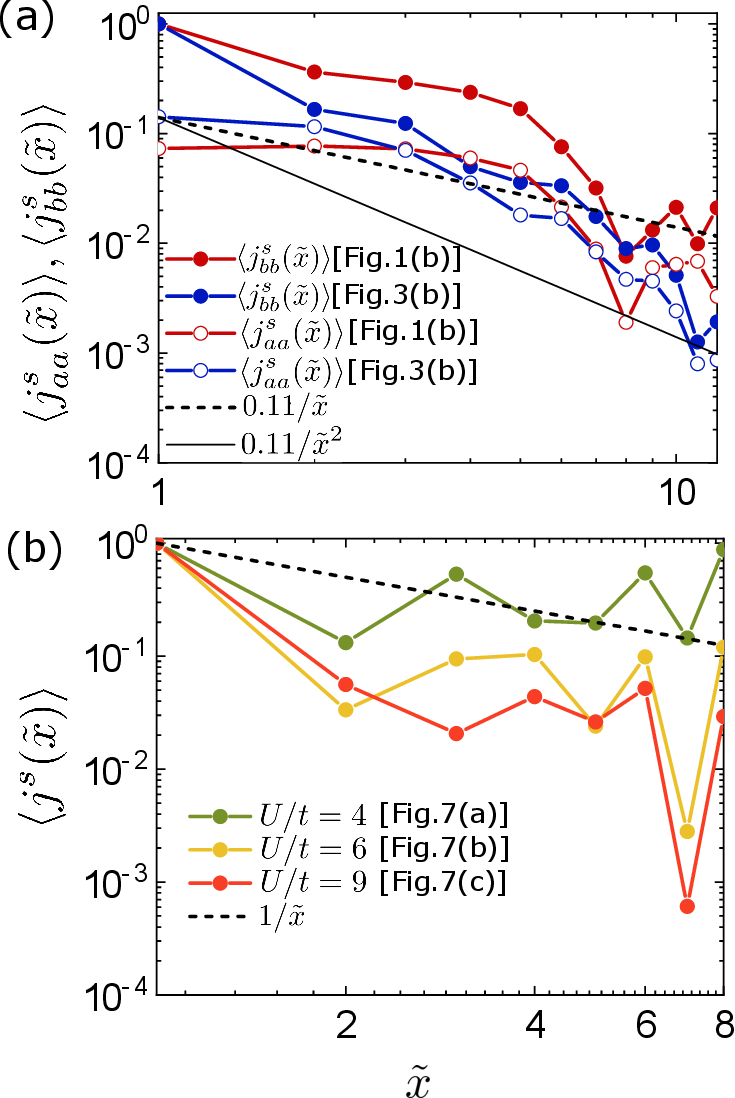}
    \caption{
    Spatial distribution of spin current as a function of the distance $\tilde x$ from the bond to which the pinning field is applied.
    (a) $\langle j^{s}_{aa} (\tilde x)\rangle$ and $\langle j^{s}_{bb} (\tilde x)\rangle$ for the carrier-doped excitonic insulators (denoted by red circles) and for the orbital-selective Mott insulators (denoted by blue circles) in the area enclosed by the blue dotted rectangles in Figs.~\ref{Fig1}(b) and \ref{Fig3}(b), respectively.    
    For comparison, power-law functions $0.11/\tilde x$ and $0.11/\tilde x^{2}$ are also plotted by the black dashed and solid lines, respectively. 
    (b) $\langle j^{s} (\tilde x)\rangle$ for the carrier-doped two-dimensional Mott insulators in the area enclosed by the blue dotted rectangles in Figs.~\ref{Fig7}(a), \ref{Fig7}(b), and \ref{Fig7}(c) for $U/t=4$, $6$, and $9$, respectively.
    $\langle j^{s} (\tilde x)\rangle$ only for nearest-neighbor bonds along the $y$ direction is shown.
    For comparison, a power function $1/\tilde x$ is also plotted by the black dashed line.
    }
    \label{Fig11}
\end{figure}

In this appendix, we demonstrate the existence of an exciton condensation in the ETHM defined by the Hamiltonian $\mathcal{H}_{\rm ETH}$ [see Eqs.~(\ref{eq-twoHub}) and (\ref{eq-crysfield})] on a ladder lattice studied in Sec.~\ref{sec3a}.
For this purpose, we evaluate the hybridization average $\langle v \rangle := \frac{1}{L}\sum_{i,\tau}\langle c_{i,a,\tau}^{\dag} c_{i,b,\tau} \rangle$ between orbitals $a$ and $b$.
Figure~\ref{Fig10} shows our DMRG results for $U=4$, $J_\text{H}=U/4$, $(t_{aa},t_{bb})=(0.4,-0.2)$, and $t_{ab}=t_{ba}=0.05$ in units of eV at electron density $n=1.92$.
These parameters are the same as those used in Fig.~\ref{Fig1}.
Our calculations clearly find that $\langle v \rangle$ is nonzero for $3\le \Delta \le 3.8$, thus including $\Delta=3$ for which the sLC textures emerge, as shown in Fig.~\ref{Fig1}(b).
\vspace{1cm}

\section{Power-law behavior of the spin current induced by a pinning field}\label{app-c}

To give further insight on the sLC textures, in this appendix, we show log-log plots of the spatial distribution of the spin current for the two-orbital Hubbard ladders in Fig.~\ref{Fig11}(a) and for the single-orbital Hubbard model on a square lattice in Fig.~\ref{Fig11}(b).
Red circles in Fig.~\ref{Fig11}(a) show the spatial distribution of $\langle j^{s}_{aa} (\tilde x)\rangle$ and $\langle j^{s}_{bb} (\tilde x)\rangle$ for the carrier-doped excitonic insulators in the area enclosed by the blue dotted rectangles in Fig.~\ref{Fig1}(b).
Here, $\tilde x=x-10.5$ is the distance from the bond to which the pinning field is applied. 
Blue circles in Fig.~\ref{Fig11}(a) show the spatial distribution of $\langle j^{s}_{aa} (\tilde x)\rangle$ and $\langle j^{s}_{bb} (\tilde x)\rangle$ for the carrier-doped orbital-selective Mott insulators in the area enclosed by the blue dotted rectangles in Fig.~\ref{Fig3}(b). 
Apart from the vicinity of the boundaries of the systems, $\langle j^{s}_{aa} (\tilde x)\rangle$ and $\langle j^{s}_{bb} (\tilde x)\rangle$ in Fig.~\ref{Fig11}(a) appear to closely follow a power-law behavior, indicated by the black dashed and solid lines.
The power-law decay seems to have the form of $\tilde x^{-l}$ with $1\lesssim l\lesssim 2$.

Figure~\ref{Fig11}(b) shows the spatial distribution of $\langle j^{s} (\tilde x)\rangle$ for the carrier-doped two-dimensional Mott insulators in the area enclosed by the blue dotted rectangles in Figs.~\ref{Fig7}(a), \ref{Fig7}(b), and \ref{Fig7}(c) for $U/t=4$, $6$, and $9$, respectively.
Here, $\langle j^{s} (\tilde x)\rangle$ is plotted only for nearest-neighbor bonds along the $y$ direction.
$\tilde x=x+1$ is the distance from the bond to which the pinning field is applied.
Because of the small cluster size, these results provide only limited information, but we find that $\langle j^{s} (\tilde x)\rangle$ follows a power-law behavior $\tilde x^{-l}$ with the exponent $l$ that becomes closer to 1 as $U/t$ approaches 4.

\end{document}